\documentclass{article}

\usepackage[english]{babel}
\usepackage[utf8]{inputenc}

\usepackage{bm}
\usepackage{graphicx}
\usepackage{amsmath}
\usepackage{hyperref}

\usepackage{authblk}

\def\LL{\widehat{L}}
\def\rmd{\mathrm{d}}
\def\rme{\mathrm{e}}
\def\rmi{\mathrm{i}}
\def\lmbd{\bm{\lambda}}
\def\psii{\bm\psi}
\def\f{\hat{f}}
\def\g{\hat{g}}
\def\h{\hat{h}}
\def\P{\bm{P}_\sigma}

\def\F{\bm{F}}
\def\S{\bm{S}}
\def\W{\bm{W}}
\def\C{\bm{C}}

\def\T{\mathrm{T}}
\def\unity{1}
\def\Re{\mathrm{Re}}
\def\Im{\mathrm{Im}}
\def\w{\omega}

\newcommand\eref[1]{\eqref{#1}}
\newcommand\Eref[1]{Equation~\ref{#1}}
\newcommand\fref[1]{Figure~\ref{#1}}
\newcommand\sref[1]{Section~\ref{#1}}
\newcommand\Aref[1]{Appendix~\ref{#1}}

\newcommand\phsintgrnd[1][z]{q(#1,\lmbd)}
\newcommand\predexp[1][z]{q(#1,\lmbd)^{-1/2}}
\newcommand\phsintgrl[3][z]{\int_{#2}^{#3} \phsintgrnd[#1] \rmd #1}

\title{Generalized symmetry relations for connection matrices in the phase-integral method}

\author[1,2]{A.~G. Kutlin \thanks{anton.kutlin@gmail.com}}
\affil[1]{Institute of Applied Physics of Russian Academy of Sciences, 46 Ulyanov str., 603950 Nizhny Novgorod, Russia}
\affil[2]{Max-Planck-Institut f\"ur Physik komplexer Systeme, N\"othnitzer Stra{\ss}e~38, 01187-Dresden, Germany}

\begin{document}



\maketitle

\begin{abstract}
We consider the phase-integral method applied to
an arbitrary ordinary linear differential equation of the second-order and study 
how its symmetries affect the connection matrices associated with its general solution.
We reduce the obtained exact general relation for the matrices to its limiting case introducing
a concept of the effective Stokes constant. We also propose a concept of an effective Stokes 
diagram which can be a useful tool for analyzing difficult equations. We show that effective 
Stokes domains which can be overlapped by a symmetry transformation are associated with the same 
effective Stokes constant and can be described by the same analytical function. Basing on
the derived symmetry relations, we propose a way to write functional equations for 
the effective Stokes constants. Finally, we provide a generalization of the derived symmetry 
relations for an arbitrary order linear system of the ordinary linear differential equations. 
This work also contains an example of usage of the presented ideas in a case of a real physical problem. To access the HTML version of the paper \& discuss it with the author, visit \url{https://enabla.com/pub/1108}.
\end{abstract}


\section{Introduction \label{sec:intro}}
Consider an arbitrary ordinary linear second-order differential equation 
written in the form of a stationary one-dimensional Schr\"odinger equation
\begin{equation}
\LL(z,\lmbd)y(z,\lmbd)=0, \quad \LL(z,\lmbd)=\frac{\rmd^2}{\rmd z^2} + R(z,\lmbd),   \label{eq:gen}
\end{equation}
where $\lmbd$ is a set of the problem's parameters. Its approximate local solution can be
obtained with use of the phase-integral approximation generated 
from the unspecified base function \cite{frbook}:
\begin{subequations}
\label{eq:phsint}
\begin{align}
y(z) &\sim c_+y_+(z) + c_-y_-(z), \label{eq:gensol}
\\
y_\pm(z) &= \predexp \exp [\pm \rmi \w(z)], \label{eq:phbase}
\\
\w(z)&=\phsintgrl[\xi]{(z_0)}{z}, \label{eq:phase}
\end{align}
\end{subequations}
where the explicit form of $q(z)$ depends on the order and particular type of the approximation.
The simplest and the best-known type is one of the WKBJ \cite{wkb1,wkb2,wkb3,wkbj}; 
it takes the form \eref{eq:phsint} with $\phsintgrnd = \sqrt{R(z,\lmbd)}$. 

The function $\w(z)$ is the phase integral, and therefore we call $\phsintgrnd[\xi]$ 
the phase integrand. Also, we will refer to $z_0$ as a basepoint.
A meaning of the brackets in the lower limit of integration is 
a bit tricky; such a notation was introduced by Fr\"oman and Fr\"oman \cite{frpaper} 
to make the integral look similar for all orders of approximation. In the lowest order and, 
particularly, in case of the WKBJ approximation, this integral is just a usual integral from $z_0$ to $z$.

The general solution \eref{eq:gensol} is a local, not global, solution of \eref{eq:gen}, i.e.
the coefficients $c_\pm$ vary from one point of the complex plane to another. Provided
that
\begin{eqnarray}
\varepsilon = q^{-3/2} \rmd^2 q^{-1/2}/\rmd z^2  + (R - q^2)/q \ll 1   \label{eq:cond}
\end{eqnarray}
in the considered area of the complex plane, the variations are, in general, slow
but may have abrupt changes on the so-called Stokes lines 
(Stokes phenomenon \cite{stokes,rwbook,heading,frbook}). Such 
changes have a form of a single-parameter linear transform \cite{heading}; 
the parameter is called the Stokes constant. 
Knowing all Stokes constants associated with a particular 
equation makes it possible to obtain a globally defined approximate solution 
of \eref{eq:gen}, and the phase-integral method provides a simple 
way to do it \cite{heading,rwbook}. Unfortunately, there are very few 
cases when this method allows finding all the constants exactly, so approximations of different 
kinds are commonly used \cite{rwbook,ours}. This happens mostly because of a lack of the exact 
equations for the Stokes constants.

The phase-integral approximation has an extensive application in various fields of physics.
It is widely used in quantum mechanics and nuclear physics to calculate energy spectrum and 
study wave functions for both  
Schr\"odinger \cite{serg96,serg02,aleixo00} 
and Dirac \cite{esp09} equations. 
It is successfully used in plasma physics to study electromagnetic waves' 
scattering characteristics \cite{gosp17, kut17}
as long as many problems in this area can be reduced to a problem of 
linear coupling \cite{shal08,shal10,shal12}. 
It is also useful in other branches of physics such as 
high-energy physics \cite{poor16} 
or general relativity \cite{ander92,manor77,rojas07}.
Among the publications devoted to the phase-integral method there are both 
classical monographs \cite{dingle73,berry91,rwbook,heading,frbook} 
and modern attempts to improve its accuracy \cite{ours,delabaere97}. 
In the present paper we provide a way to reduce the number 
of independent Stokes constants with use of the symmetries of \eref{eq:gen}. 

The paper is organized as follows. 
In \sref{sec:fmtrintro} we recall briefly the F-matrix formalism and introduce a concept of the 
effective Stokes constant. We also give a matrix formulation of the well-known Heading's rules for
analytical continuation; such formulation appears to be much more convenient for our purposes than 
the traditional one, which can be found, for example, in \cite{rwbook}.
In \sref{sec:fmtrsymm} we derive the symmetry relations for the connection matrices in the most common 
case of the second-order differential equation.
In \sref{sec:scsymm} we simplify the relations by reducing F-matrix to its limiting form and
obtain the symmetry relations for the effective Stokes constants. In this section we also discuss
a possibility of writing functional equations for effective Stokes constants.
In \sref{sec:effsd} a concept of an effective Stokes diagram is presented; this concept is a natural
consequence of the effective Stokes constants' usage.
In \sref{sec:weber} we present an example of a real physical problem solved with the help of our technique; 
the symmetry relations for effective Stokes constants are used to find the exact 
form of reflection and transmission coefficients for the Weber equation. 
In \sref{sec:discuss} we discuss an applicability of the symmetry relations derived in \sref{sec:fmtrsymm}
for the case of an arbitrary order system of ordinary linear differential equations.
And, finally, in \sref{sec:cnclsns} are the conclusions.

\section{Connection matrices and effective Stokes constants \label{sec:fmtrintro}}
As it was mentioned in the previous section, the coefficients $c_\pm$ 
vary from one point of the complex plane to another. Taking into consideration
the linearity of \eref{eq:gen}, these variations can by described formally by means 
of the F-matrices \cite{frbook}:
\begin{eqnarray}
\psii(z_2) = \F(z_2,z_1) \psii(z_1),
\label{eq:fmtrdef}
\end{eqnarray}
where $\psii(z) = [{c_+(z),c_-(z)}]^{\T}$ and `T' denotes the transpose operation.
In principle, F-matrix can be obtained exactly from the corresponding differential 
equation \cite{frbook}; then, the approximate solution \eref{eq:gensol} 
with $c_\pm=c_\pm(z)$ becomes exact\footnote{
The functions $c_\pm(z)$ can be determined for any exact solution $y(z)$ of \eqref{eq:gen} and any phase-integral base functions $y_\pm(z)$ unequivocally from two conditions: the first one is simply $y(z)=c_+(z)y_+(z) + c_-(z)y_-(z)$, and the second one reads as $c_+'(z)y_+(z) + c_-'(z)y_-(z)=0$; this last condition allows to write first derivative of $y(z)$ formally as if $c_\pm(z)$ would be constant.
}.  

F-matrix does not depend on the $\psii-$vectors, i.e. on the initial/boundary conditions 
of \eref{eq:gen}, and represents properties of its general solution. 
As long as the general solution may have branch points, 
F-matrix can depend on the particular path $\gamma$ of analytical continuation; 
we will indicate the path by writing $\F[\gamma]$ instead of $\F(z_2,z_1)$, if necessary.
 
F-matrix also depends on the chosen system of approximate solutions $y_\pm$ and, 
particularly, on the chosen basepoint $z_0$. Consider equation \eref{eq:gen} and some definite
type of the phase-integral approximation with $q(z,\lmbd)$ as a phase integrand 
and $z_0$ as a basepoint. Then, consider two points $z_{1,2}(\lmbd)$ of the complex plane 
and an oriented curve $\gamma$ connecting these points; 
according to the definition \eref{eq:fmtrdef} of the F-matrix, 
$\F[\gamma]$ relates coefficients $c_\pm$ at the ends of the curve. We say 
that $\F_{q,z_0}[\gamma]$ performs an analytical continuation of the general solution of \eref{eq:gen} 
along the oriented curve $\gamma$ in the system $\{q,z_0\}$; we will indicate the
system by the corresponding subscript of the F-matrix, if necessary.

Consider equation \eref{eq:gen} and two F-matrices, $\F_{q,z_0}[\gamma]$ and $\F_{q,\tilde{z}_0}[\gamma]$, 
performing an analytical continuation of the general solution along the oriented curve $\gamma$; the matrices
represent the same analytical continuation, but written with use of different basepoints. They can be related as
\begin{equation}
\F_{q,\tilde{z}_0}[\gamma] = \W[\tilde{z}_0,z_0]\F_{q,z_0}[\gamma]\W[z_0,\tilde{z}_0],
\label{eq:bpchange} 
\end{equation}
where
\begin{equation}
\W[b,a] =  
\left(\begin{array}{*{2}{c}}
\rme^{\rmi \phsintgrl{(a)}{(b)}} & 0 \\ 0 & \rme^{-\rmi \phsintgrl{(a)}{(b)}} 
\end{array}\right);
\label{eq:W}
\end{equation}
this follows directly from \eref{eq:fmtrdef} and \eref{eq:phsint}. 

Then, consider the same differential equation, but different F-matrices $\F_{q,z_0}[\gamma]$ and $\F_{-q,z_0}[\gamma]$;
this is the simplest nontrivial example of F-matrices representing the same analytical continuation, but written
with use of different phase integrands. Accurate to insignificant constant multiplier connected with slow
$\predexp$ dependency, these two systems of base phase-integral approximate solutions differ by their order;
the difference can be naturally described by a permutation matrix $\P$:
\begin{equation}
\F_{-q,z_0}[\gamma] = \P\F_{q,z_0}[\gamma]\P^{-1}, \quad
\P =
\left(\begin{array}{*{2}{c}}
0 & 1 \\ 1 & 0 
\end{array}\right).
\label{eq:qchange} 
\end{equation}
The permutation matrix in this relation maps the system $\{q,z_0\}$ to the system $\{-q,z_0\}$, but the direction
of the mapping is insignificant in case of $2 \times 2$ matrices because then $\P^{-1} = \P$. 

In the limit \eref{eq:cond} of small epsilon, F-matrix represents the linear transformation
mentioned in \sref{sec:intro} and can be approximately expressed in terms of the corresponding
Stokes constants and phase integrals. As long as the limiting form of F-matrix is more common
and more convenient from the practical point of view, we will end the present section with 
its detailed description.

Let's start with some basic definitions. Hereinafter we will refer to a point $z$ of the complex
plane as a singular point if epsilon from \eref{eq:cond} is infinitely large in such a point; its
vicinity will be referred to as an interaction area. 
At every point $z$ of the complex plane except the singularities we will distinguish two 
orthogonal directions. Let's define the Stokes direction 
as a direction with $\Re \left[ \phsintgrnd \rmd z \right]=0$ and the anti-Stokes direction 
as a direction with $\Im \left[ \phsintgrnd \rmd z \right]=0$. We will also use a notion of 
the Stokes (anti-Stokes) field as a set of Stokes (anti-Stokes) directions for the entire 
complex plane. Following \cite{heading, rwbook}, we introduce Stokes (anti-Stokes) lines as a 
paths along Stokes (anti-Stokes) field emanating from the singularities. 
Any domain of the complex plane bounded by the Stokes (anti-Stokes) lines and containing no other 
Stokes (anti-Stokes) lines will be referred to as the anti-Stokes (Stokes) domain. Particularly, if 
$q^2(z,\lmbd) \sim z^n$ as $z$ goes to complex infinity, then there are $n+2$ Stokes 
(anti-Stokes) domains in the vicinity of the infinity -- we will call such domains the Stokes 
(anti-Stokes) wedges. Asymptotic phase-integral solutions \eref{eq:phbase} oscillate along anti-Stokes 
lines with constant flow and increase (or decrease) exponentially with constant phase along Stokes lines. 
The increasing (decreasing) solution will be called dominant (subdominant) in a given Stokes domain. 
Also, the complex plane with singular points, Stokes and anti-Stokes lines and branch cuts associated 
with a branching structure of asymptotic solutions \eref{eq:phsint} will be referred to as a Stokes diagram.

Consider a Stokes domain and the F-matrix associated with crossing this domain in 
counterclockwise direction relatively to the chosen basepoint. According to estimates 
made by Fr\"oman and Fr\"oman \cite{frbook}, in a limit \eref{eq:cond} of small epsilon 
such F-matrix can be approximately written as either $\S[s]$ or $\S^{\T}[s]$, where
\begin{equation}
\S[s] = \left(\begin{array}{*{2}{c}} 1 & 0 \\ s & 1 \end{array}\right)    
\end{equation}
and $s$ is a Stokes constant associated with the domain. 
In the present paper, we will call such a constant an \emph{effective Stokes constant} 
to emphasize its difference from the traditional one. Indeed, there can be multiple Stokes 
lines in the Stokes domain, and every such line is associated with its own traditional Stokes 
constant \cite{heading, rwbook}, whereas the effective Stokes constant is associated with 
the whole Stokes domain. It is also worth mentioning that every effective Stokes constant can be 
expressed in terms of the traditional ones; it becomes clear from the reasoning presented in \Aref{app1}.

We will refer to $\S$ as a Stokes operator. As it can be inferred from the Fr\"omans estimates,
we must use $\S$ for a Stokes domain where $y_+$ is dominant and $\S^{\T}$ otherwise. The rules
for the Stokes operator are analogous to the traditional Heading's rules for analytical 
continuation, presented, for example, in \cite{heading, rwbook}. As it follows from the previous discussion
and from the Heading's rules itself, both traditional and effective Stokes constants depend on basepoint.
In \cite{heading, rwbook} the change of the basepoint is called `reconnection', thus we will
refer to $W[b,a]$ introduced by \Eref{eq:W} as a reconnection operator.

Aside from crossing Stokes line and changing basepoint, the Heading's rules include the rule for crossing a 
branch cut associated with the branching structure of the phase-integral approximate solutions \eref{eq:phbase}.
According to Fr\"oman and Fr\"oman \cite{frbook}, the operator, describing crossing a branch cut emerging 
from the first order zero of the squared phase integrand $q^2(z,\lmbd)$ in counterclockwise direction, takes the form
\begin{equation}
\C =  \left(\begin{array}{*{2}{c}} 0 & -\rmi \\ -\rmi & 0 \end{array}\right).    
\label{eq:C}
\end{equation}
In case of the different order of the branch point the operator must be raised to an appropriate power.

With help of these three operators, $\S$, $\W$ and $\C$, the reader can perform any analytical continuation far
away from any singularities; the matrix formulation of the traditional Heading's rules appears
to be much more convenient for our purposes.

\section{Symmetry relations for the connection matrices \label{sec:fmtrsymm}}

Let's clarify what we mean by a symmetry of a given equation. Here we introduce three operators
$\f(z,\lmbd)$, $\g(z,\lmbd)$ and $\h(z,\lmbd)$ such that
\begin{equation}
\f:\{y\} \rightarrow \{y\}, \quad
\g:\{z\} \rightarrow \{z\}, \quad
\h:\{\lmbd\} \rightarrow \{\lmbd\},
\end{equation}
i.e. $\f$ (complex conjugation, multiplication, etc.) acts on the set of functions, preserving 
the structure of their Stokes diagrams, $\g$ (complex conjugation, invertible function of 
complex variable, etc.) acts on the field of complex numbers, and $\h$ (complex conjugation, 
analytical function of $\lmbd$, etc.) acts on the parameter's space. As it will be seen from the
following discussion, the restriction on $\g$ to be invertible is crucial. 

We define any set of operators $\{\f(z,\lmbd),\g(z,\lmbd),\h(z,\lmbd)\}$ as a symmetry 
transformation of \eref{eq:gen}
if for any its solution $y(z,\lmbd)$ there is another solution $\tilde{y}(z,\lmbd)$ such that
$\tilde{y}(z,\lmbd)=\f(z,\lmbd)y(\g(z,\lmbd)z,\h(z,\lmbd)\lmbd)$, i.e.
\begin{equation}
\LL(z,\lmbd)y(z,\lmbd) \equiv 0 \ \Longrightarrow\  
\LL(z,\lmbd) \left[ \f(z,\lmbd)y(\g(z,\lmbd)z,\h(z,\lmbd)\lmbd) \right] \equiv 0.   \label{eq:symdef}
\end{equation}
As it follows from the definition, 
$\tilde{y}(z,\lmbd)$ as well as $y(z,\lmbd)$ can be written in the form \eref{eq:gensol},
but with different coefficients $c_\pm$ and $\tilde{c}_\pm$ correspondingly. 
This difference and, as a consequence, restrictions on the connection matrices 
can be obtained directly by applying the symmetry transformation to \eref{eq:phbase} as it was 
done in \cite{frsymm} for the special case analysed by Fr\"omans, but we prefer more intuitive derivation.

Let's choose some definite type of the phase-integral approximation with $q(z,\lmbd)$ 
as the phase integrand; this phase integrand will be used all throughout the present section. 
Consider $y(z,\lmbd)$, some solution of \eref{eq:gen}, and its analytical continuation
along the oriented curve $\gamma$, performed in the system $\{q,z_0\}$:
\begin{equation}
y(z,\lmbd):\ \gamma\ \longrightarrow\ \F_{q,z_0}[\gamma,\lmbd]. 
\end{equation}
Then, consider another solution of \eref{eq:gen}, $\tilde{y}(z,\lmbd)$, and its
analytical continuation along the same curve, but performed with use of different basepoint:
\begin{equation}
\tilde{y}(z,\lmbd):\ \gamma\ \longrightarrow\ \F_{q,\tilde{z}_0}[\gamma,\lmbd]. 
\end{equation}
As it follows from the previous discussion, $\F_{q,z_0}[\gamma,\lmbd]$ and 
$\F_{q,\tilde{z}_0}[\gamma,\lmbd]$ represent the same operation in different bases and
can be related with use of a reconnection operator. On the other hand, and this is much more important, 
if the solution $\tilde{y}(z,\lmbd)$ is connected with $y(z,\lmbd)$ through
the symmetry transformation $\{\f(z,\lmbd),\g(z,\lmbd),\h(z,\lmbd)\}$, its analytical continuation
along the curve $\gamma$ can be obtained directly from the 
corresponding continuation of $y(z,\lmbd)$ by a formal application of the symmetry transformation:
\begin{equation}
\tilde{y}(z,\lmbd) \equiv \f y(\g z,\h \lmbd):\ 
\gamma\ \longrightarrow\ \f\F_{q,z_0}[\g\gamma,\h\lmbd]. 
\end{equation}
We underline that this last analytical continuation is performed along the same curve $\gamma$ as it was in 
the previous cases because any curve is the ordered set of allowed values of $z$, not $\g z$. Thus, 
$\f\F_{q,z_0}[\g\gamma,\h\lmbd]$ and $\F_{q,\tilde{z}_0}[\gamma,\lmbd]$ represent the 
same analytical continuation, but performed with use of different systems of base approximate phase-integral 
solutions \eref{eq:phbase}. It is crucial that the systems can differ not only by the basepoints' location,
but also by the functions $y_\pm$. Indeed, accurate to insignificant constant multiplier connected with slow
$\predexp$ dependency, there are two possible mappings implemented by the symmetry transformation: 
$y_\pm \rightarrow y_\pm$, or $y_\pm \rightarrow y_\mp$; such mappings were considered in the previous section 
and can be described by a permutation matrix $\P$.
 
Now, we are ready to write the final relation for the F-matrices. A formula expressing this relation is shown below:
\begin{align}
\label{eq:fmtrsymm}
&\P^{-1} \left\{ \f \F_{q,z_0} \left[\g \gamma, \h \lmbd \right] \right\} \P\\
\notag &\qquad =
   \W \left[ \g^{-1} z_0(\h\lmbd), \tilde{z}_0(\lmbd) \right]
   \F_{q,\tilde{z}_0} \left[ \gamma, \lmbd \right]
   \W \left[ \tilde{z}_0(\lmbd), \g^{-1} z_0(\h\lmbd) \right].
\end{align}
The inverse operator $\g^{-1}$ in the formula appears due to the variable\linebreak exchange in the phase integral:
\begin{equation}
\g \int_{(z_0)}^z q(\xi) \rmd \xi = \int_{(z_0)}^{\g z} q(\xi) \rmd \xi 
= \int_{(\g^{-1}z_0)}^{z} q(\g\xi) \rmd \g\xi.
\end{equation}
An explicit form of the permutation matrix $\P$ can be determined by a direct application of the symmetry 
transformation to the base functions $y_\pm$.

The relation \eref{eq:fmtrsymm} has a fairly transparent structure. Indeed, assume we chose $y_\pm$ 
as a base system of phase-integral approximate solutions with $\g^{-1} z_0(\h\lmbd)$ as a basepoint. 
To implement the analytical continuation in such a base with use of known F-matrices 
$\f\F_{q,z_0}[\g\gamma,\h\lmbd]$ and $\F_{q,\tilde{z}_0}[\gamma,\lmbd]$, 
we have to either change the basepoint from $\g^{-1} z_0(\h\lmbd)$ to $\tilde{z}_0$ and apply
$\F_{q,\tilde{z}_0}[\gamma,\lmbd]$ (right hand side of \eref{eq:fmtrsymm}) 
or choose an appropriate order of the base functions $y_\pm$ and apply $\f\F_{q,z_0}[\g\gamma,\h\lmbd]$
(left hand side of \eref{eq:fmtrsymm}); our relation reflects the equivalence of these two approaches.

\begin{figure}[ht]
\centering
\noindent
\includegraphics[width=\textwidth]{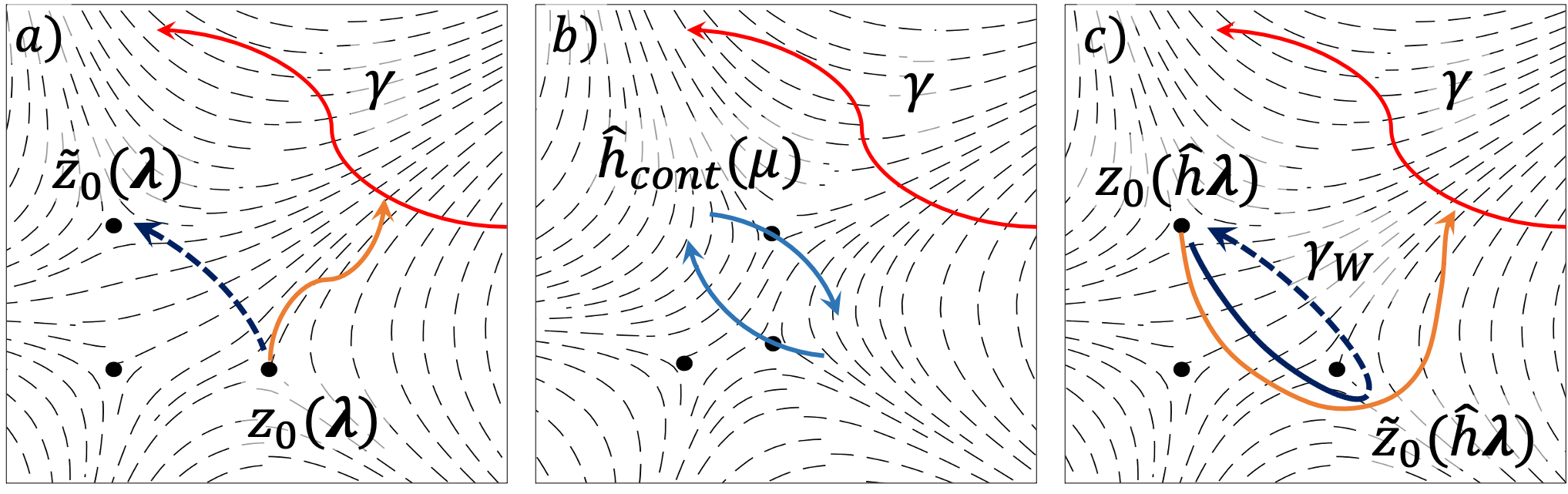}
\caption{
Stokes field for $q(z)=\sqrt{z(z-1)(z-\rmi)}$ and its metamorphosis under the action of $\hat{h}$ swapping singular points in $z=1$ and $z=\rmi$ as shown in the panel (b) (${\hat{f}=\hat{g}=1}$). This is an example of how to determine $\gamma_W$, the integration path for the reconnection operator $\W \left[\g^{-1} z_0(\h\lmbd), \tilde{z}_0(\lmbd) \right]$ from \eqref{eq:fmtrsymm}. In the panel (a) the curve $\gamma$ (continuous red) and two different basepoints are shown, $z_0(\lmbd)$ and $\tilde{z}_0(\lmbd)$; the continuous orange line illustrates the path used for the definition of the phase integral $\omega(z)$ from \eqref{eq:phsint}, and the dashed dark blue line shows a path we want to use for the reconnection from $z_0$ to $\tilde{z}_0$.
In the panel (b), some intermediate state of the continuously parameterized symmetry transformation $\hat{h}_{cont}(\mu)$ is shown. Finally, in the right panel (c) the result of the transformation and the path $\gamma_W$ are shown; the first part of the path $\gamma_W$ (solid dark blue line) corresponds to the symmetry transformation itself, and the second part (dashed dark blue line) corresponds to the reconnection path shown in the panel (a).}
\label{fig:recon}
\end{figure}  

A presence of the reconnection operator $\W \left[\g^{-1} z_0(\h\lmbd), \tilde{z}_0(\lmbd) \right]$ in \eqref{eq:fmtrsymm} implies an integration along some path $\gamma_W$ in the complex plane;
thus, the resulting relation may depend on this path if the phase integrand has poles
or branch points. This fact forces us to consider not only the final result of the 
symmetry transformation, but even how exactly it was performed; i.e., we have to parameterize
$\g$ and $\h$ operators. This can be done by introducing of an auxiliary variable $\mu$ varying 
from $0$ to $1$ such that, for example, $\g_{cont}(\mu)=\unity+\mu (\g-\unity)$, and $\g_{cont}(1)=\g$. 
Then, the path $\gamma_W$ can be determined in the following way: one have to fix the integration path used in the phase-integral $\omega(z)$ from \eqref{eq:phsint} for the basepoint $z_0(\lmbd)$ (continuous orange line in Fig.~\ref{fig:recon}(a)),
study how this path deforms under our transformation, 
go along this deformed path from $\hat{g}^{-1}z_0(\hat{h}\lmbd)$ to $z_0(\lmbd)$, and then perform just a usual reconnection from $z_0(\lmbd)$ to $\tilde{z}_0(\lmbd)$ as if there was no symmetry transformation involved. Note that $\hat{g}^{-1}z_0(\hat{h}\lmbd)=\tilde{z}_0(\lmbd)$ doesn't in general mean that the reconnection operator is the identity operator; in principle, $\gamma_W$ can form closed loops like the one in Fig.~\ref{fig:recon}.

The relation \eref{eq:fmtrsymm} is actually a generalization of the symmetry relations obtained by
Fr\"oman and Fr\"oman \cite{frsymm}. It takes the form presented in their article
with $\{\f = \g = \h = c.c.\}$, where $c.c.$ stands for the complex conjugation, for the
first case they considered, and with $\{\g = -\unity; \f = \h = \unity\}$ for the second case.

\section{Symmetry relations and functional equations\\ for effective Stokes constants \label{sec:scsymm}}

The symmetry relation \eref{eq:fmtrsymm} is exact and quite general, but it is not very useful
from the practical point of view because F-matrices are usually used in a limit \eref{eq:cond} of
small epsilon; thus, analogous symmetry relations for the Stokes constants appear to be more
convenient. In the present section, we obtain the symmetry relations for the effective Stokes constants 
and discuss the consequences of the relations.

\begin{figure}
\centering
\noindent
\includegraphics[width=\textwidth]{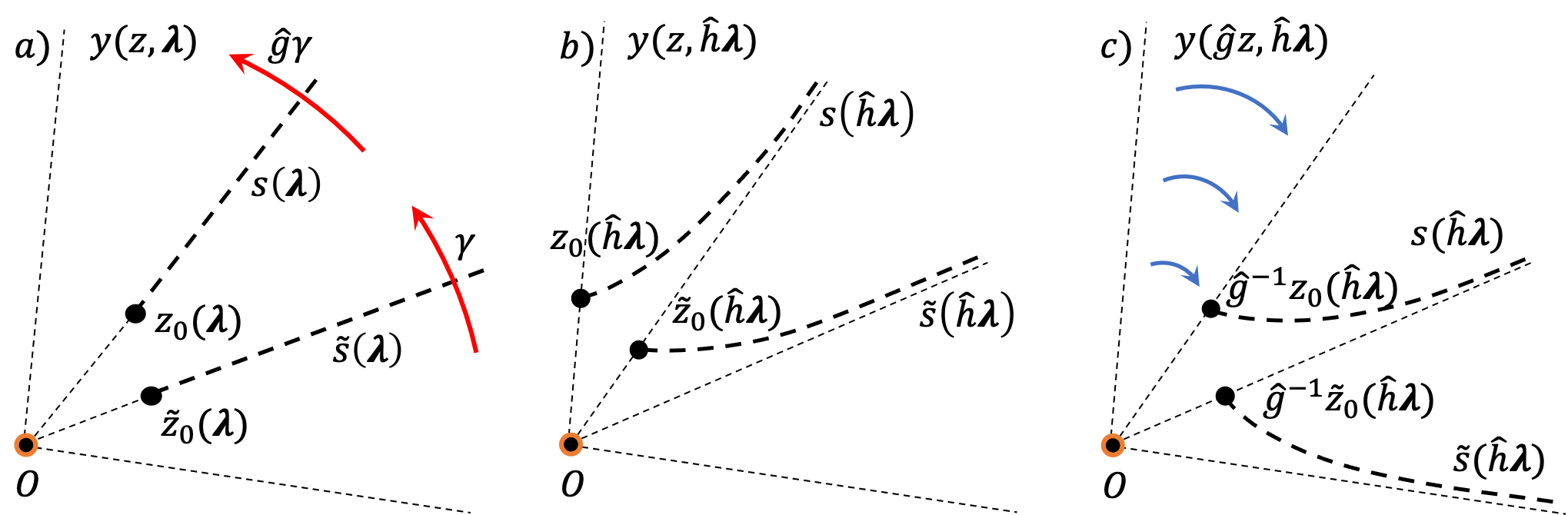}
\caption{An illustration for Eq.~\ref{eq:gensym} with $\hat{f}=1$; the point $O$ is a center of the rotation $\hat{g}$, the thin straight dashed lines emanating from $O$ represent the Stokes directions, the bold dashed lines emanating from the base points stand for the Stokes lines, and $\hat{g}^{-1}z_0(\hat{h}\lmbd)=z_0(\lmbd)$. Here you can see how the Stokes constants $s$ and $\tilde{s}$ from different Stokes directions can be related due to the overlap generated by the symmetry transformation: in the left panel (a) the two paths $\gamma$ and $\hat{g}\gamma$ are shown; in the middle panel (b) only the parameter transformation $\hat{h}$ is applied moving the basepoints $z_0$ and $\tilde{z}_0$, and in the right panel (c) the rotation $\hat{g}$ finishes the transformation overlapping the Stokes direction corresponding to $s(\hat{h}\lmbd)$ with the one which previously corresponded to $\tilde{s}(\lmbd)$.}
\label{fig:exmpl1}
\end{figure}  

Consider an oriented curve $\gamma$ located far away from any interaction area and crossing
one Stokes domain associated with the effective Stokes constant $\tilde{s}$ in counterclockwise direction 
relative to the chosen basepoint as shown in Fig.~\ref{fig:exmpl1}. Consider another Stokes domain associated with
another effective Stokes constant $s$ such that the transformed oriented curve $\g\gamma$ also crosses this second 
domain far from any singularity. 
Using the relation \eref{eq:fmtrsymm} for the F-matrices, its limiting form described in the 
\sref{sec:fmtrintro}, and the rules for the Stokes operators, we obtain
\begin{align}
\label{eq:gensym}
\S^{(\T)} \left[ \f s(\h\lmbd) \right] &= 
\W \left[ \g^{-1}z_0(\h\lmbd), \tilde{z}_0(\lmbd) \right]\\
\notag &\quad\times\S^{(\T)} \left[ \tilde{s}(\lmbd) \right]
\W \left[ \tilde{z}_0(\lmbd), \g^{-1}z_0(\h\lmbd) \right],
\end{align}
where $s$ and $\tilde{s}$ are the effective Stokes constants having $z_0$ and $\tilde{z}_0$ as their basepoints, and $\S^{(\T)}$ can be either $\S$ or $\S^{\T}$. The choice in the right hand side 
of \eref{eq:gensym} must be made on the basis of general rules from \sref{sec:fmtrintro}: 
we use $\S$ for the Stokes domain associated with $\tilde{s}(\lmbd)$ if $y_+$ is dominant 
and $\S^{\T}$ otherwise. In the left hand side of \eref{eq:gensym} we have to use the 
same form as in the right hand side; the rule is a consequence of the permutation
matrix presence in \eref{eq:fmtrsymm}.

The relation obtained above allows us to take a fresh look at the nature of the 
different effective Stokes constants. Two different Stokes domains which can be overlapped by the 
transformation $\g$ are actually associated with the same effective Stokes constant but 
evaluated in the different points of the parameters' space. In particular, every set of Stokes 
wedges can be described by a single multidimensional analytical complex function $s(\lmbd)$.

\begin{figure}
\centering
\noindent
\includegraphics[width=\textwidth]{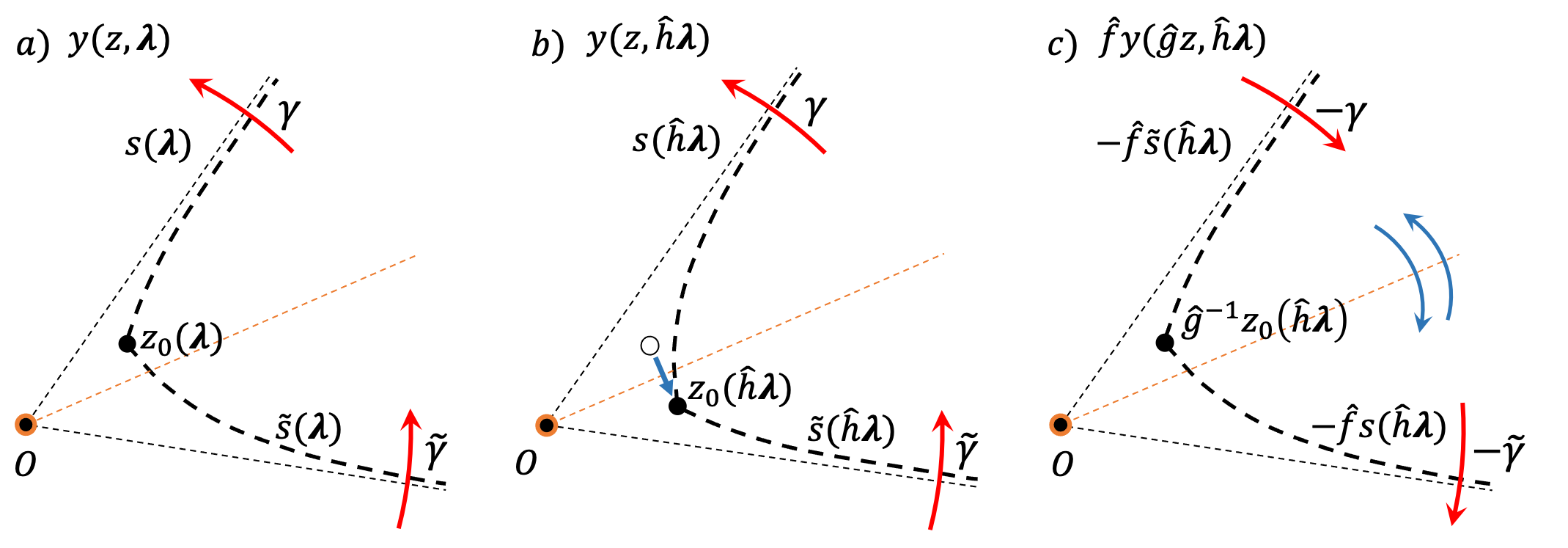}
\caption{An illustration for the ``conjugation" symmetry for which the complex conjugation symmetry discussed around the Eq.~\ref{eq:cnjgtn} stands as a special case. The orange dashed line starting at the point $O$ represents the conjugation axis, the thin straight dashed lines emanating from $O$ represent the Stokes directions, and the bold dashed lines emanating from the basepoint $z_0$ stand for the Stokes lines; for the simplicity we assume here that $z_0\equiv \tilde{z}_0$. You can see how the Stokes constant $s$ can be related to its ``mirror image" $\tilde{s}$ with minus sign: the ``conjugation" applied between the steps (b) and (c) not only overlapped different Stokes directions but also inverted the directions on the oriented curves $\gamma$ and $\tilde{\gamma}$.}
\label{fig:exmpl2}
\end{figure}

Consider an important special case of a complex conjugation symmetry, 
analysed by Fr\"omans \cite{frsymm}. 
For every equation which is real on the real axis the symmetry $\f=\g=\h=c.c.$ holds, 
i.e. $\LL^*(z^*,\lmbd^*)=\LL(z,\lmbd)$ and
\begin{equation}
\S^{(\T)} \left[ - s^*(\lmbd^*) \right] = 
\W \left[ z_0^*(\lmbd^*), \tilde{z}_0(\lmbd) \right]
\S^{(\T)} \left[ \tilde{s}(\lmbd) \right]
\W \left[ \tilde{z}_0(\lmbd), z_0^*(\lmbd^*) \right].
\label{eq:cnjgtn}
\end{equation}
A minus sign before the effective Stokes constant in the left hand side of \eref{eq:cnjgtn} is a result 
of the complex conjugation symmetry; it appears for any transformation $\g$ which changes the direction
of analytical continuation relative to the chosen basepoint since the Stokes constant is defined for the 
counterclockwise direction, see Fig.~\ref{fig:exmpl2}.
If $\tilde{z}_0(\lmbd) = z_0^*(\lmbd^*)$ and $\lmbd$ is real, we obtain a result 
$s^*(\lmbd) = - \tilde{s}(\lmbd)$, which can be inferred from the symmetry relations presented in \cite{frsymm}. 
If furthermore $s = \tilde{s}$, we get an important and extremely simple relation $ s^*(\lmbd) = - s(\lmbd)$, 
i.e. such effective Stokes constant is purely imaginary provided $\lmbd$ is real.

Another important case of \eref{eq:gensym} is a case with $\g=\unity$ and $z_0(\lmbd) = \tilde{z}_0(\lmbd)$:
\begin{equation}
\S^{(\T)} \left[ \f s(\h\lmbd) \right] = 
\W \left[ z_0(\h\lmbd), z_0(\lmbd) \right]
\S^{(\T)} \left[ s(\lmbd) \right]
\W \left[ z_0(\lmbd), z_0(\h\lmbd) \right].
\label{eq:func}
\end{equation}
Such case is relevant for every equation. The obtained relation is a functional 
equation for the considered effective Stokes constant. Usually such equation helps to illuminate 
a branching structure of the effective Stokes constant and write it as a new single-valued function 
multiplied by the known multivalued one. It is worth mentioning that even simple formal 
symmetries like $\{\h=\rme^{2\rmi\pi},\g=\f=\unity\}$ may produce nontrivial functional equations; 
it happens every time when the reconnection operator differs from unity.

\section{Effective Stokes diagram \label{sec:effsd}}

\begin{figure}
\centering
\noindent
\includegraphics[width=\textwidth]{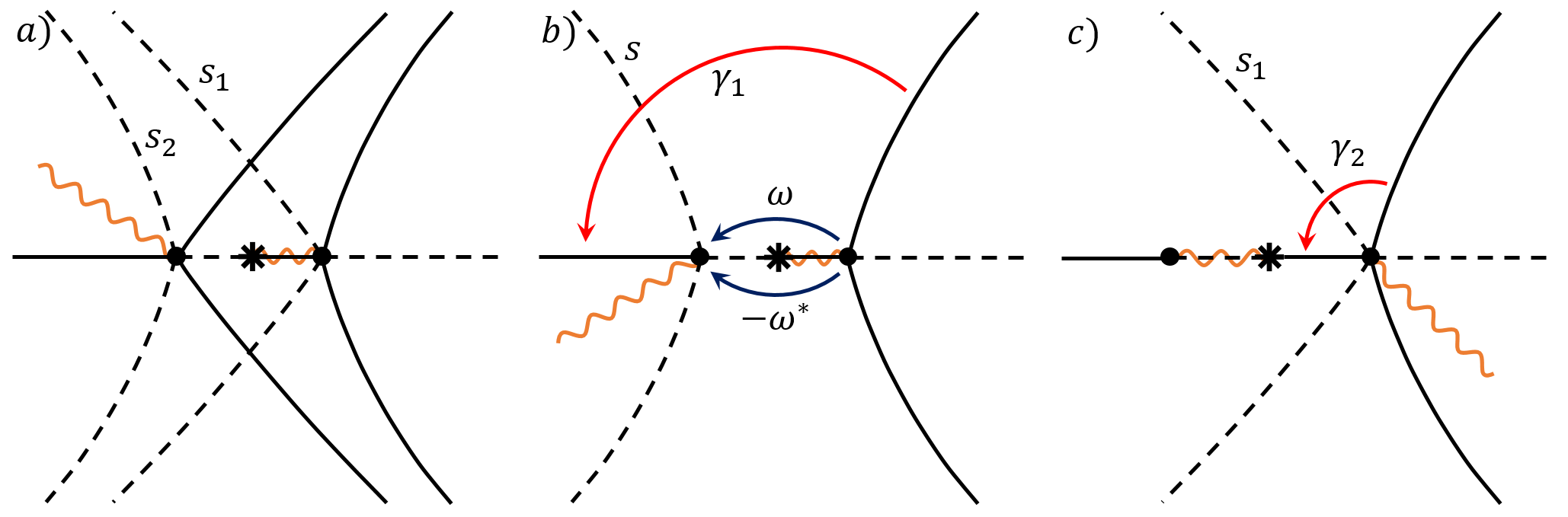}
\caption{
Stokes diagrams for $q(z,g)=\sqrt{-z + g^2/z}$; Stokes lines are dashed. 
(a) A traditional Stokes diagram; dots and stars indicate correspondingly zeros and poles of the phase integrand.
(b) An effective Stokes diagram with a specified path of analytical continuation $\gamma_1$. 
(c) An effective Stokes diagram with a different path of analytical continuation $\gamma_2$.
The Stokes constant $s_1$ is meaningful only if $g \gg 1$ and all the singularities are far away 
from each other -- otherwise the phase-integral approximation itself fails between 
the pole and the right zero and the exact F-matrix must be used.}
\label{fig:effsd_1}
\end{figure}  

\begin{figure}
\centering
\noindent
\includegraphics[width=\textwidth]{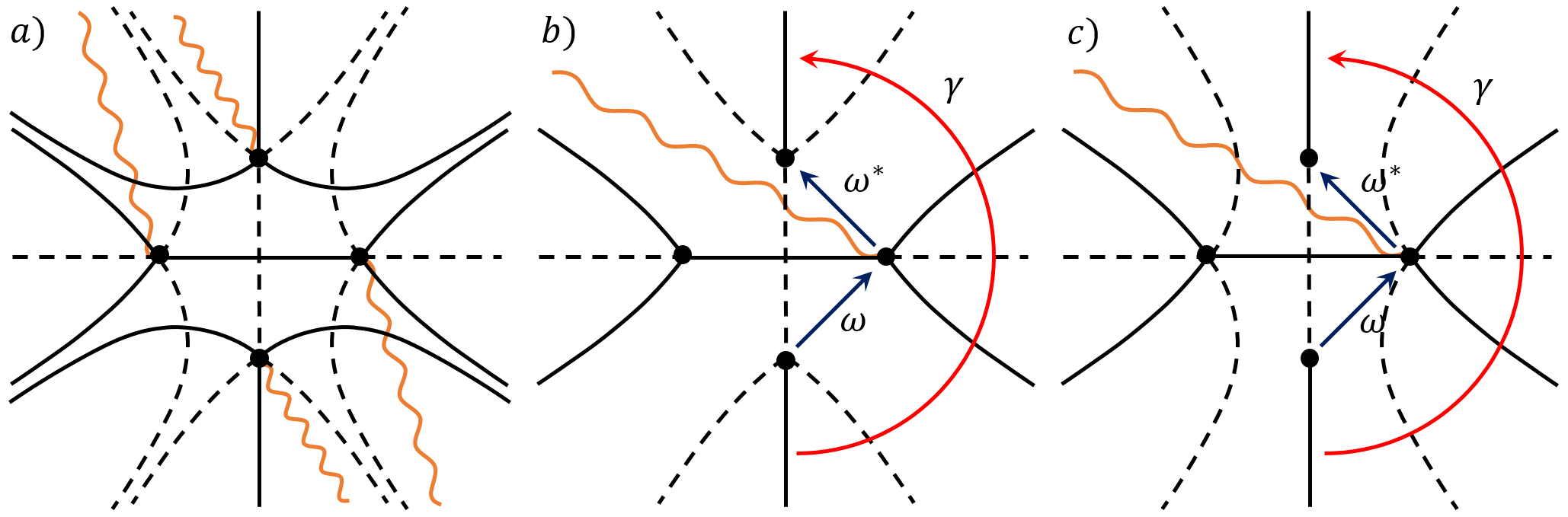}
\caption{
Stokes diagrams for $q(z,E)=\sqrt{z^4-E}$; Stokes lines are dashed.
(a) A traditional Stokes diagram; dots mark the zeros of the phase integrand.
(b) An effective Stokes diagram with a specified path of analytical continuation $\gamma$.
(c) An effective Stokes diagram with the same path of analytical continuation but different basepoints.}
\label{fig:effsd_2}
\end{figure}

The concept of the effective Stokes constant allows us to introduce a notion of the effective Stokes line.
Since every Stokes domain can be described by a single effective Stokes constant, we can visualize this fact 
by plotting a single effective Stokes line instead of a set of traditional Stokes lines. 
Similarly, every set of anti-Stokes lines located in the same anti-Stokes
domain can be replaced by a single effective anti-Stokes line; such a line now is just a borderline 
separating different Stokes domains. A Stokes diagram consisting of effective Stokes and anti-Stokes 
lines will be referred to as an effective Stokes diagram. As it can be seen from Figures \ref{fig:effsd_1} 
and \ref{fig:effsd_2}, effective diagram is not unique. 
It can be plotted differently depending on a chosen path of analytical 
continuation (\fref{fig:effsd_1}); that is why it is usually convenient to specify 
the path right on the diagram. But, even when the path is chosen, we
are still free to vary our basepoints' locations. Actually, the effective Stokes line can connect any two 
points of the corresponding Stokes domain if it does not destroy the topology of the entire 
Stokes diagram taking into account the specifics of a particular problem. 
Choosing what points to connect by the effective Stokes line we will indicate what basepoints 
will be used in the given Stokes domain (\fref{fig:effsd_2}). Also it can be useful to indicate 
values of the phase integrals instead of plotting multiple branch cuts.

The main purpose of the effective Stokes diagram is to indicate clearly and transparently what
basepoint is used to cross a given Stokes domain. For some values of parameters $\lmbd$, such
diagram can be similar to the traditional one, but for other values the diagrams will differ
for sure because effective Stokes line is always emerge from the same basepoint $z_0(\lmbd)$
while a traditional one follows the Stokes field. Moreover, effective Stokes diagram
is less detailed than the traditional one and may be more convenient for the analysis of 
complicated equations.

\section{Example: the Weber equation \label{sec:weber}}

\begin{figure}
\centering
\noindent
\includegraphics[width=\textwidth]{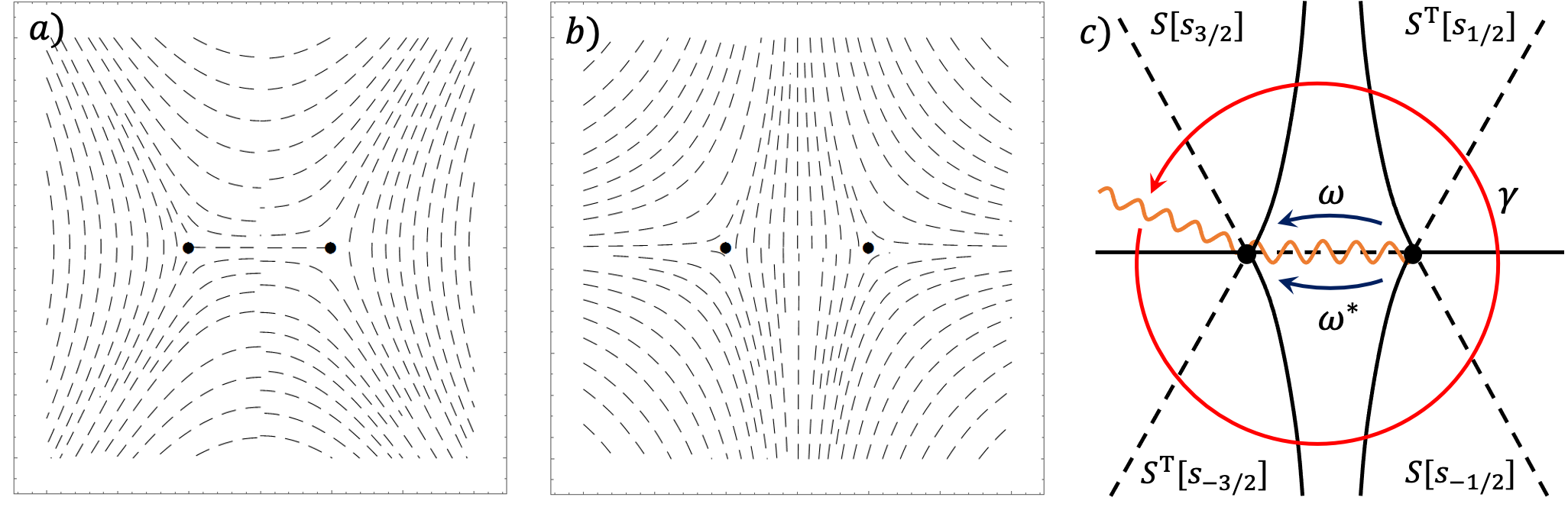}
\caption{Stokes field (a), anti-Stokes field (b) and an effective Stokes diagram (c) 
for the Weber equation \eref{eq:weber}.}
\label{fig:wsd}
\end{figure} 

We will use WKBJ approximation all throughout this section. Also in this section we will use the symbol 
$\W[\w] \equiv \W \left[\phsintgrl{a}{b} \right] \equiv \W[b,a]$ to underline a value of the phase integral.

Now let us consider the Weber equation
\begin{equation}
\frac{\rmd^2 y(z,\delta)}{\rmd z^2}+(z^2-\delta^2)y(z,\delta)=0
\label{eq:weber}
\end{equation}
with the boundary conditions of a presence of incident wave from the large negative $z$ and an absence of such 
a wave from the large positive $z$. We define the reflection (transmission) coefficient $\rm{\mathcal{R}}$ 
($\rm{\mathcal{T}}$) as a ratio of the amplitudes of the reflected (transmitted) and incident waves. 
Our aim now is to find these scattering characteristics.

The boundary conditions can be written in terms of $\psii$-vectors as
\begin{equation}
\psii_0 = \left(\begin{array}{*{2}{c}} 1 \\ 0 \end{array}\right),
\label{eq:wbound}
\end{equation}
where $\psii_0$ is a $\psii$-vector in an anti-Stokes wedge containing a ray $Arg(z)=0$.
First of all, scattering characteristics must be written through the 
Stokes constants. To do this, we must analytically continue our boundary 
condition \eref{eq:wbound} to $z$ with $Arg(z)=\pi$. 
Using \fref{fig:wsd} and the rules from \sref{sec:fmtrintro}, we write
\begin{equation}
\psii_{\pi} = 
\S \left[ s_{3/2} \right]
\W \left[ \w(\delta) \right] 
\S^{\T} \left[ s_{1/2} \right] \psii_0 \equiv 
\rme^{\rmi \w(\delta)} \left(\begin{array}{*{2}{c}} 1 \\ s_{3/2} \end{array}\right),
\end{equation}
where $\w(\delta)=-\rmi\pi\delta^2/2$ is a phase integral calculated above the cut 
from $z=\delta$ to $z=-\delta$. Now we have to identify incident, reflected and transmitted waves. 
Since $y_+ \propto e^{\rmi z^2/2}$ hence it is 
an outgoing wave for $z \rightarrow +\infty$ as well as for $z \rightarrow -\infty$ and
\begin{equation}
\mathcal{R} = \frac{1}{s_{3/2}},\quad \mathcal{T} = \rmi\frac{\rme^{-\rmi w}}{s_{3/2}}.
\label{eq:RT}
\end{equation}

To find $s_{3/2}$, let's try a traditional method described, for example, in \cite{frpaper, rwbook}. 
We can obtain the desired equations for the Stokes constants using a single-valuedness of the general 
solution and analytically continuing it around the origin far away from the interaction area along the
oriented curve $\gamma$ (\fref{fig:wsd}(c)):
\begin{equation}
\unity = 
\C^2
\S \left[ s_{3/2} \right]
\W \left[ \w \right] 
\S^{\T} \left[ s_{1/2} \right]
\S \left[ s_{-1/2} \right]
\W \left[ \w \right]
\S^{\T} \left[ s_{-3/2} \right],
\end{equation}
from which follows
\begin{equation}
\begin{cases}
s_{1/2} = s_{-3/2}\\
s_{3/2} = s_{-1/2}\\ 
s_{1/2}s_{3/2} + e^{-2 \rmi w} + 1 = 0.
\label{eq:webtrad}
\end{cases}
\end{equation}
The $\C$ operator is squared here because the squared phase integrand $R(z,\lmbd)$ 
is asymptotic to $z^2$ as $z$ goes to complex infinity.

As we can see from the system \eref{eq:webtrad}, we cannot find $s_{3/2}$ using the traditional method -- 
we need at least one more restriction for the Stokes constants. In \cite{rwbook} a 
requirement of the flux conservation is used as the restriction 
and it gives $s_{-1/2}=-s_{1/2}^*$ -- it helps to determine the absolute value of the
reflection coefficient, but its phase stays unknown. This example shows that even in 
such a simple situation as the Weber equation, the traditional approach cannot fully 
resolve the problem.
 
Actually the flux conservation is a consequence of the real-valuedness and regularity of the coefficients 
of the Weber equation and hence the same relation can be obtained from the 
complex conjugation symmetry using \eref{eq:cnjgtn}. Moreover, first two equations from \eref{eq:webtrad} 
are just a consequence of a symmetry $\{\g=\rme^{\rmi\pi},\h=\f=\unity\}$ -- and it is clear 
because the symmetry is just a rotation and can be seen even from the usual analytical continuation. 
Therefore, the only original equation in the system \eref{eq:webtrad} is the last one -- it cannot 
be obtained from any other considerations. But these are not the only consequences of the 
equation's symmetries, so let's write them all.

The first nontrivial symmetry relation can be obtained from the symmetry 
$\{\g=\h=\rmi,\f=\unity\}$. It can overlap, for example, Stokes domains associated with
$s_{1/2}$ and $s_{-1/2}$. Both of the Stokes constants have the same 
basepoint $z_0(\delta)=\delta$ and $\g^{-1}z_0(\h\delta)=(-\rmi)\rmi\delta=\delta$, so
\begin{equation}
\S \left[ s_{1/2}(\rmi \delta) \right] = 
\W \left[ \delta, \delta \right]
\S \left[ s_{-1/2}(\delta) \right]
\W \left[ \delta, \delta \right],
\end{equation}
or
\begin{equation}
s_{1/2}(\rmi \delta) = s_{-1/2}(\delta).
\label{eq:websym_1}
\end{equation}
Considering \eref{eq:webtrad}, we can 
see now that all four Stokes wedges can be described by only one function as it
was mentioned in the \sref{sec:scsymm}. 

\begin{figure}
\centering
\noindent
\includegraphics[width=\textwidth]{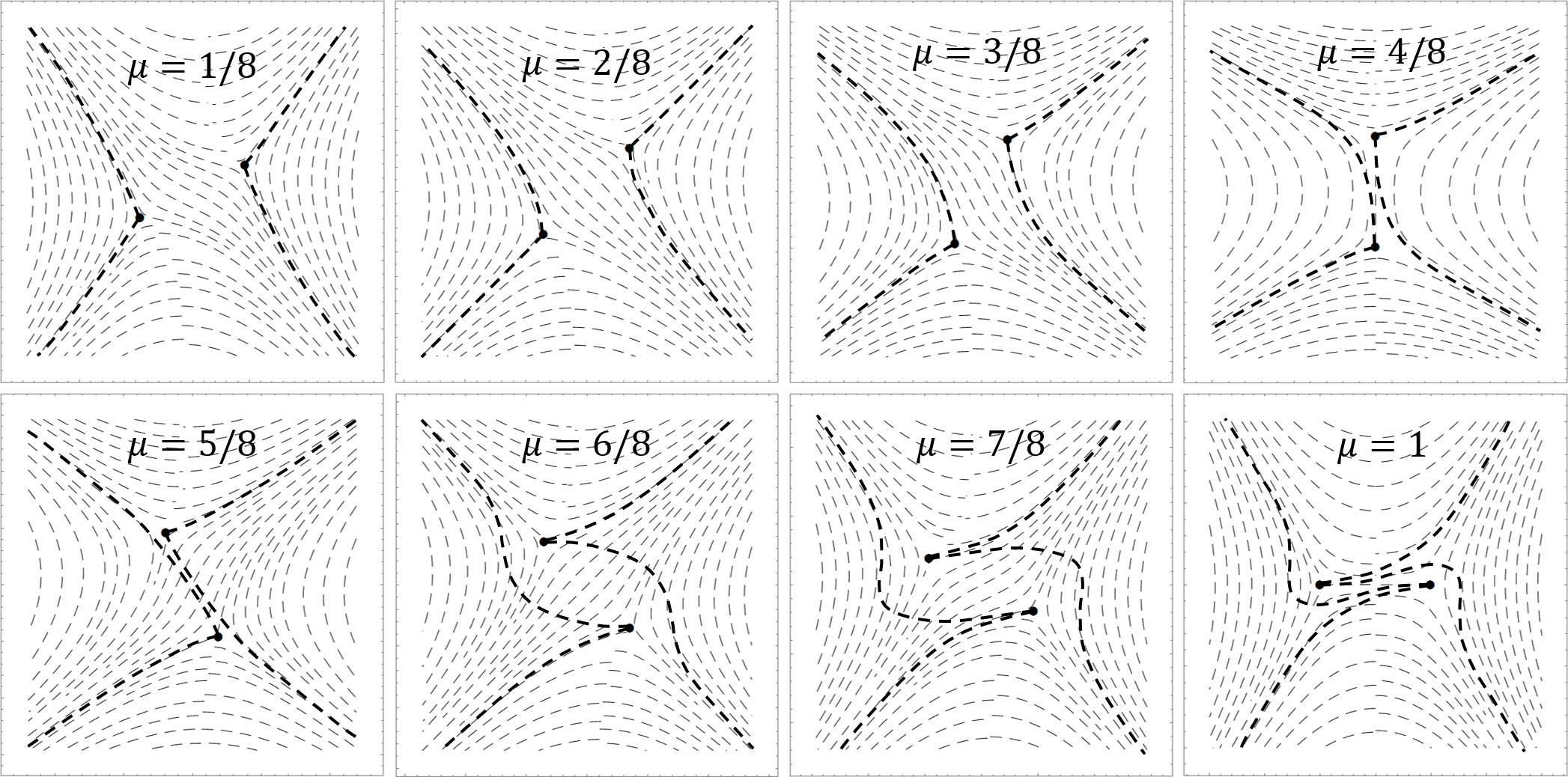}
\caption{Evolution of the Stokes field and effective Stokes lines 
under the continuous parameter's transformation $\h_{cont}(\mu)=\rme^{\rmi\pi\mu}$.}
\label{fig:webrs}
\end{figure}

Now consider another symmetry $\{\h=\rme^{\rmi\pi},\g=\f=\unity\}$. 
As it was written in the \sref{sec:scsymm}, such a symmetry gives rise to a functional 
equation, which can help to illuminate a branching structure of the Stokes constant. 
For definiteness, we will talk about $s_{3/2}$. To understand what to choose as endpoints 
in the phase integrals in \eref{eq:func}, let's parametrize $\h$ as $\h_{cont}(\mu)=\rme^{\rmi\pi\mu}$ 
and look at \fref{fig:webrs}. For $s_{3/2}$, the basepoint $z_0(\delta)=\delta \rme^{\rmi\pi}$, 
and the question is how it is changing under our transformation. Using \fref{fig:webrs} 
we can see that finally it arrives at the point $z=\delta$, but the phase difference between the
initial and the final positions matters because it defines an integration path.
That is why we have to write $z_0(\h\delta)=\delta \rme^{2\rmi\pi}$ and
\begin{equation}
\S \left[ s_{3/2}(\delta \rme^{\rmi\pi}) \right] = 
\W \left[- \w \right]
\S \left[ s_{3/2} (\delta) \right]
\W \left[  \w \right],
\end{equation}
or
\begin{equation}
s_{3/2}(\delta \rme^{\rmi\pi})=s_{3/2}(\delta)\rme^{-2\rmi \w}=s_{3/2}(\delta)e^{-\pi\delta^2}.
\label{eq:websym_2}
\end{equation}
This functional equation can be solved by substitution 
$s_{3/2}(\delta)=\rmi\delta^{\rmi\delta^2}f(\delta^2)$, where $f(\delta^2)$
is single-valued in a sense $f(x)=f(x \rme^{2\rmi\pi})$.

And, finally, consider a conjugation symmetry $\f=\g=\h=c.c.$. This symmetry relates the 
Stokes constants in the upper half of the complex plane to the constants in the lower half. 
In particular, according to \eref{eq:cnjgtn},
\begin{equation}
\S \left[ -s_{1/2}^*(\delta^*) \right] = 
\W \left[ z_0^*(\delta^*), \tilde{z}_0(\delta) \right]
\S \left[ s_{-1/2}(\delta) \right]
\W \left[ \tilde{z}_0(\delta),z_0^*(\delta^*) \right],
\end{equation}
and, since $z_0(\delta)=\tilde{z}_0(\delta)=\delta$,
\begin{equation}
s_{1/2}^*(\delta^*)=-s_{-1/2}(\delta).
\label{eq:websym_3}
\end{equation}
For real values of $\delta$ the last relation is nothing but the law of the flux conservation 
mentioned above. But, for complex values, together with \eref{eq:websym_2} and \eref{eq:webtrad} 
it gives
\begin{equation}
s_{3/2}(\delta)=s_{-1/2}(\delta)=\rmi(\rmi \delta^2)^{\rmi \delta^2/2}p(\rmi \delta^2),
\label{eq:websym_4}
\end{equation}
where $p(x)=p(x \rme^{2\rmi\pi})$ and $p(x)$ is real on the real axis. Now, using \eref{eq:websym_1} 
and the last equation from the system \eref{eq:webtrad}, we can write a functional equation for $p(x)$:
\begin{equation}
p(x)p(-x)=2\cos(\pi x/2).
\label{eq:pfunc}
\end{equation}
This functional equation is similar to the Euler's reflection formula \cite{gamma} and can be reduced 
to the formula by substitutions $p(x)=u(x)\sqrt{2\pi}/\Gamma(1/2+x/2)$ and $x=1-2t$, where $u(x)u(-x)=1$. 
The function $u(x)$ can be found from the boundary conditions of \eref{eq:pfunc}. Indeed, we know 
exactly \cite{rwbook} that $s_{3/2}(0)=\rmi\sqrt{2}$. We also can assume, according to the approximation of 
isolated singularities \cite{rwbook,ours}, that every Stokes constant approaches an imaginary 
unit as $\delta$ goes to plus infinity. Taking into consideration \eref{eq:websym_4}, we can say that
\begin{equation}
\begin{cases} 
p(0) = \sqrt{2} \\
p(x) \sim x^{-x/2}\ as\ x \rightarrow \pm \rmi \infty. 
\end{cases}
\end{equation}
Using the asymptotics of gamma function and definition of $u(x)$, we can finally write 
that $u(x)=(2 \rme)^{-x/2}$ and
\begin{equation}
s_{3/2}(\delta)=\rmi(\rmi\delta^2)^{\rmi\delta^2/2}
\frac{\sqrt{2\pi}(2\rme)^{-\rmi\delta^2/2}}{\Gamma(1/2+\rmi\delta^2/2)}.
\label{eq:s3/2}
\end{equation}
This is an exact expression for the effective Stokes constant for 
the Weber problem -- it can be verified using 
an exact solution of \eref{eq:weber} as it was done in \cite{ours}. 
Now the desired scattering characteristics \eref{eq:RT} can be found.

\section{Discussion on the possible generalizations \label{sec:discuss}}

The symmetry relation \eref{eq:fmtrsymm} for connection matrices presented in \sref{sec:fmtrsymm} 
was obtained for the case of the second-order differential equation, but appears to be much more general. 
In the present section we discuss the range of its applicability.

Consider an arbitrary order system of the first order linear ordinary differential equations:
\begin{equation}
\frac{\rmd}{\rmd z} \bm{y}(z,\lmbd) = \bm{M}(z,\lmbd) \bm{y}(z,\lmbd),
\label{eq:mgen}
\end{equation}
where $\bm{y}$ is a vector of functions to determine and $\bm{M}(z,\lmbd)$ is a square matrix of corresponding dimension.
Assume we can write an approximate phase-integral local solution of \eref{eq:mgen} in the way
similar to the case of the second-order equation \eref{eq:gen}:
\begin{subequations}
\label{eq:mphsint}
\begin{align}
\bm{y}(z,\lmbd)& \sim \bm{c} \cdot \bm{\tilde{y}}(z,\lmbd), \label{eq:mgensol}
\\
\bm{\tilde{y}}_i(z,\lmbd) &= \bm{A}_i(z,\lmbd) \exp [\rmi \w_i(z,\lmbd)], \label{eq:mphbase}
\\
\w_i(z,\lmbd)&=\int_{(z_0)}^z q_i(z,\lmbd) \rmd z, \label{eq:mphase}
\end{align}
\end{subequations}
where $\bm{A}_i$ is a polarization vector, $\bm{c}$ is a vector of coefficients (analogue of $c_\pm$), 
$\bm{\tilde{y}}$ is a vector of approximate phase-integral solutions \eref{eq:mphbase} (analogue of $y_\pm$), 
and '$\cdot$' stands for their inner product.
Then, introducing F-matrix and defining symmetry completely analogous to \sref{sec:fmtrintro} and \sref{sec:fmtrsymm},
we can exploit exactly the same reasoning and arrive to exactly the same symmetry relation with the only one difference:
aside from changing basepoint and reordering of the set of the base functions, we have to multiply each function $\bm{\tilde{y}}_i$ 
by an appropriate constant $a_i$; the constant was insignificant for the previous discussion because $y_\pm$ from \eref{eq:phbase} 
would have been multiplied by the same constant $a = a_+ = a_-$. In the general case considered here the multiplication
must be described by a diagonal matrix $\bm{\mathit{\Lambda}}$ such that $\bm{\mathit{\Lambda}}_{ii} = a_i$, i.e. the generalization of the
symmetry relation \eref{eq:fmtrsymm} takes the form
\begin{align}
\label{eq:mfmtrsymm}
&\bm{\mathit{\Lambda}}^{-1}\P^{-1} \left\{ \f \F_{\bm{q},z_0} \left[\g \gamma, \h \lmbd \right] \right\} \P \bm{\mathit{\Lambda}} \\
 \notag &\qquad=  \W \left[ \g^{-1} z_0(\h\lmbd), \tilde{z}_0(\lmbd) \right]
   \F_{\bm{q},\tilde{z}_0} \left[ \gamma, \lmbd \right]
   \W \left[ \tilde{z}_0(\lmbd), \g^{-1} z_0(\h\lmbd) \right],
\end{align}
where $\bm{q}$ is a vector of the phase integrands $q_i$ and $\W$ has an appropriate dimension.
The diagonal elements of the matrix $\bm{\mathit{\Lambda}}$, as well as an explicit form of the permutation matrix $\P$, can
be determined by a direct application of the symmetry transformation to the base functions $\bm{\tilde{y}}_i(z,\lmbd)$.

\section{Conclusion \label{sec:cnclsns}}

The method of phase integrals is a beautiful and powerful method of a linear ordinary 
differential equations' asymptotic analysis, but its range of applicability is highly 
restricted to relatively simple problems; more complicated problems need additional equations 
for the Stokes constants. The analysis presented in this paper allows the reader to find 
functional relations between connection matrices and thus to reduce the number of unknowns.

The main result of the present work is stated by \Eref{eq:mfmtrsymm}. This result, 
like any other obtained in the paper, is valid for any approximation of the 
phase-integral type, not only for WKBJ approximation. Furthermore, this result holds 
not only for the second-order, but also for the arbitrary order system of linear ordinary 
differential equations. 

The symmetry relation \eref{eq:mfmtrsymm} and its two-dimensional version \eref{eq:fmtrsymm} are quite general, 
but not very useful from the practical point of view. To overcome the difficulty, we rewrote the relations in the
most common case of the second-order differential equation with use of the limiting form of 
F-matrix and the concept of effective Stokes constant. The rewritten
symmetry relation is stated by \Eref{eq:gensym}; we also introduced a concept of effective Stokes diagram
which can be a useful tool for the analysis of complicated equations. 

We showed that Stokes domains which can be overlapped by the variable transformation $\g$ are actually 
associated with the same effective Stokes constant and can be described by the same analytical function. 
We showed that every differential equation has some formal symmetries (e.g. $\{\h=\rme^{2\rmi\pi},\g=\f=\unity\}$) 
which may lead to nontrivial relations for the Stokes constants. We also showed that the symmetry relations allow the 
reader to write functional equations for the effective Stokes constants; such functional equations
help to illuminate a branching structure of the effective Stokes constant and write it as a new single-valued 
function multiplied by the known multivalued one.

The functional relations which can be obtained from \Eref{eq:mfmtrsymm} are likely to be as 
complex as the initial differential equation; however, they are strict and can be used as 
a basis for the construction of perturbation theory (will be a matter of a separate paper).

\subsection*{Acknowledgments}
This work was supported by the Russian Science Foundation (grant No~14-12-01007). The author is grateful to
Dr. A.~G.~Shalashov for careful reading of the manuscript and useful comments.

\appendix

\section{Connection between traditional and \\ effective Stokes constants \label{app1}}

Imagine a Stokes domain with several traditional Stokes lines 
emerging from the common basepoint. Crossing this domain in terms of the traditional Stokes constants
implies multiple sequential applications of either $\S$ or $\S^{\T}$ operators. 
However, as we can see by a direct calculation, Stokes operators $\S$ as well as $\S^{\T}$ 
form a multiplicative group:
\begin{equation}
\S[s_2]\S[s_1] = \S[s_2 + s_1], \quad
\S^{\T}[s_2]\S^{\T}[s_1] = \S^{\T}[s_2 + s_1].
\end{equation}
Consequently, every such set of traditional Stokes constants can be replaced by a single one;
the single Stokes constant is just a sum of the traditional constants.

Now consider the reconnection operator $\W$. As it follows from the properties of the phase integral, 
these operators also form a multiplicative group:
\begin{equation}
\W[c,b]\W[b,a] = \W[c,a].
\end{equation}
And, completely analogous to the situation discussed above, every set of sequential changes of the basepoint
can be described by a single reconnection operator.

Finally, imagine a Stokes domain with multiple Stokes lines and multiple basepoints. Crossing this 
domain in terms of $\S$ and $\W$ operators looks like
\begin{equation}
\S[s_n]\W[a_n,a_{n-1}]\S[s_{n-1}]\W[a_{n-1},a_{n-2}]\ ...\ \S[s_1]\W[a_1,a_0]\S[s_0].
\end{equation}
Note that for any $s'$ there is such $s''$ that $\S[s']\W[b,a]=\W[b,a]\S[s'']$, therefore
we can change the order of $\S$ and $\W$ operators. Hence,
every such set of operators can always be replaced by a single combination $\S[s_l] \W[a_n,a_0]$ or, 
if someone prefers different ordering, $\W[a_n,a_0] \S[s_r]$. Consequently, every Stokes domain can 
be described by the effective Stokes constant $s$ despite the number of Stokes lines it contains.


\end{document}